\documentclass{article}

\usepackage{arxiv}
\usepackage{amsthm, amsmath, amsbsy, amssymb, booktabs, epsfig, epsf}
\usepackage{multicol, multirow, psfrag, booktabs, graphicx, subfig, rotating}
\usepackage[authoryear]{natbib}
\usepackage[colorlinks, citecolor = blue, urlcolor = blue]{hyperref}

\newtheorem{proposition}{Proposition}

\newcommand{\PC}[1]{\ensuremath{\left(#1\right)}}
\newcommand{\chav}[1]{\ensuremath{\left\{#1\right\}}}
\newcommand{\bm}[1]{{\boldsymbol {#1}}}

\title{Non-Separable Spatio-temporal Models via Transformed Gaussian Markov Random Fields}

\author{
  Douglas  R. M. Azevedo \\
  Department of Statistics\\
  Universidade Federal de Minas Gerais\\
  Av. Presidente Ant\^{o}nio Carlos 6627 \\
  Pampulha Belo Horizonte, Minas Gerais \\
  31270-901, Brazil\\
  \texttt{douglasrm.azevedo@gmail.com} \\
  \And
  Marcos O. Prates \\
  Department of Statistics\\
  Universidade Federal de Minas Gerais\\
  Av. Presidente Ant\^{o}nio Carlos 6627 \\
  Pampulha Belo Horizonte, Minas Gerais \\
  31270-901, Brazil\\
  \texttt{marcosop@est.ufmg.br} \\
  \And
  Michael R. Willig \\
  Department of Ecology \& Evolutionary Biology \\ 
  University of Connecticut \\
  75 N. Eagleville Rd. U-3043\\
  Storrs, Connecticut 06269, U.S.A.\\
  \texttt{michael.willig@uconn.edu} \\
}

\begin{document}
\maketitle

\begin{abstract}
Models that capture the spatial and temporal dynamics are applicable in many science fields. Non-separable spatio-temporal models were introduced in the literature to capture these features. However, these models are generally complicated in construction and interpretation. We introduce a class of non-separable Transformed Gaussian Markov Random Fields (TGMRF) in which the dependence structure is flexible and facilitates simple interpretations concerning spatial, temporal and spatio-temporal parameters. Moreover, TGMRF models have the advantage of allowing specialists to define any desired marginal distribution in model construction without suffering from spatio-temporal confounding. Consequently, the use of spatio-temporal models under the TGMRF framework leads to a new class of general models, such as spatio-temporal Gamma random fields, that can be directly used to model Poisson intensity for space-time data. The proposed model was applied to identify important environmental characteristics that affect variation in the abundance of \textit{Nenia tridens}, a dominant species of snail in a well-studied tropical ecosystem, and to characterize its spatial and temporal trends, which are particularly critical during the Anthropocene, an epoch of time characterized by human-induced environmental change associated with climate and land use.
\end{abstract}


\section{Introduction} \label{sec:introduction}

In many fields of science, spatio-temporal models are useful to better understand and more realistically represent the dynamics of systems of interest.  This is particularly true for ecological systems during the Anthropocene \citep{steffen2007anthropocene, zalasiewicz2010new}, time of rapid, human-induced environmental change linked to climate and land use. Ecological systems (suites of species that co-occur in time and space, and that interact with each other, as well as with matter and energy, to form systems) are complex, involving dynamics associated with abiotic (e.g., temperature, precipitation) and biotic (e.g., land use composition and configuration) characteristics.  Because the Anthropocene is characterized by unprecedented rates of change, it is important to understand and predict spatio-temporal dynamics of populations that can inform management and policy with the ultimate goal of reducing the likelihood of species extinction and consequent loss of ecosystem services that are essential for human well being. The urgency of the situation is reflected in recent suggestions that the planet is now entering its sixth major extinction period as well as in the controversy surrounding the announcement of biological armageddon \citep{lister2018climate,lister2019reply,schowalter2019warnings,willig2019populations}.

Generalized Linear Mixed Models \citep[GLMM,][]{breslow1993approximate} represent a flexible class of models that are capable of accommodating random effects simply. In this class of models, it is necessary to choose an appropriate link function to model the conditional mean with covariates and random effects. Transformed Gaussian Markov Random Fields \citep[TGMRF,][]{prates2015transformed} appear as an effective tool for modeling spatial data. In this approach, it is possible to directly choose the distribution of the conditional mean, including, covariates, and to define the desired spatial structure. Unlike traditional GLMMs it is not necessary to define an appropriate link function, which in some models, can be difficult to interpret. Moreover, TGMRF's do not suffer from spatial confounding \citep{reich2006effects, hodges2010adding, hughes2013dimension, hanks2015restricted, thaden2018structural, prates2018alleviating} because of its copula-based structure \citep{hughes2015copcar, prates2015transformed}.

A simple way to include spatial dependence in statistical models is to use spatially structured random effects. For areal data, the most common spatial structure is the Conditional Autoregressive model \citep[CAR,][]{besag1974spatial}. CAR models are useful for fitting spatial data but their structure is not directly applied to multivariate problems. Multivariate Conditional Autoregressive models \citep[MCAR,][]{gelfand2003proper, carlin2003hierarchical, jin2005generalized, jin2007order} were proposed to extend CAR models when multiple variables are observed in the same space. The idea is to control for the correlation structure between variables. \cite{sain2011spatial} and \cite{rodrigues:2012} presented an alternative way to define the cross-correlation between regions and variables.

In this paper, we propose a non-separable, flexible and interpretable multivariate dependence structure and an extension of TGMRFs to multivariate problems. In particular, we define a new model from a spatio-temporal perspective. This new formulation allows a clear and direct interpretation of the contributions of spatial, temporal and spatio-temporal components. In addition, the proposed model prevents spatio-temporal confounding via a copula structure that guarantees the separation of fixed and random effects by construction. This is a clear advantage because, to the best of our knowledge, the literature does not consider how spatio-temporal random effects might confound fixed effects estimates, and no
solutions to this problem have been proposed.

We leverage a long-term (17 years) ecological study \citep{bloch2006context, willig1998long, willig2007cross, willig2014experimental} to illustrate the utility of our multivariate TGMRF approach. More specifically, we construct and interpret spatio-temporal models for counts of \textit{Nenia tridens}, an abundant species of snails, that dominates the gastropod fauna in forests of Puerto Rico. This is particularly relevant because these ecosystems are disturbance-mediated: the mapping of environmental characteristics onto geographic space changes over time in response to climatic events (e.g., cyclonic storms and droughts) and subsequent secondary succession, with consequences to the abundance and distribution of resident species. Fortunately, spatially explicit data are available for counts of species as well as for habitat characteristics that are known to influence abundance over time.

Section~\ref{sec:gastropds_abundance} highlights the ecological relevance and importance of the data. Section~\ref{sec:dependence_structure} summarizes several multivariate dependence structures in the literature as well as introducing a new proposal. Moreover, the utility of the new proposal is emphasized within an integrated discussion of existing formulations. The TGMRF formulation for the spatio-temporal setting and how inference is performed are presented in Section~\ref{sec:TGMRF}. A detailed simulation study about the proposed method is presented in Section~\ref{sec:simulation}. Section~\ref{sec:application} revisits the ecological application showing the empirical and modeled results. A final conclusion and discussion are presented in Section~\ref{sec:final_remarks}.

\section{Ecological characteristics} \label{sec:gastropds_abundance}



Gastropods (snails and slugs) are the second-most species-rich group of animals in the world \citep{prie:2019}. They are ubiquitous heterotrophs (decomposers) and provide essential ecosystem functions associated with energy flow and nutrient cycling \citep{mason1970snail,prather2013invertebrates}. Previous research has documented their habitat associations and responses to disturbances such as tree-fall gaps \citep{alvarez1993effects}, hurricanes \citep{willig1991effect, secrest1996legacy, prates2011intervention}, and previous land-use history \citep{willig1998long} in the Luquillo Experimental Forest of Puerto Rico. Thus gastropods represent an ideal taxonomy considered as an illustrative case for modeling spatio-temporal demographics in a changing environmental context. Moreover, {\it Nenia tridens} is one of the most numerically dominant gastropods on the  Luquillo Forest Dynamics Plot (LFDP), and has a heterogeneous spatial distribution, making it of particular ecological importance \citep{willig1998long,bloch2006context}.

\begin{figure}
    \centering
    \includegraphics[width=0.80\textwidth]{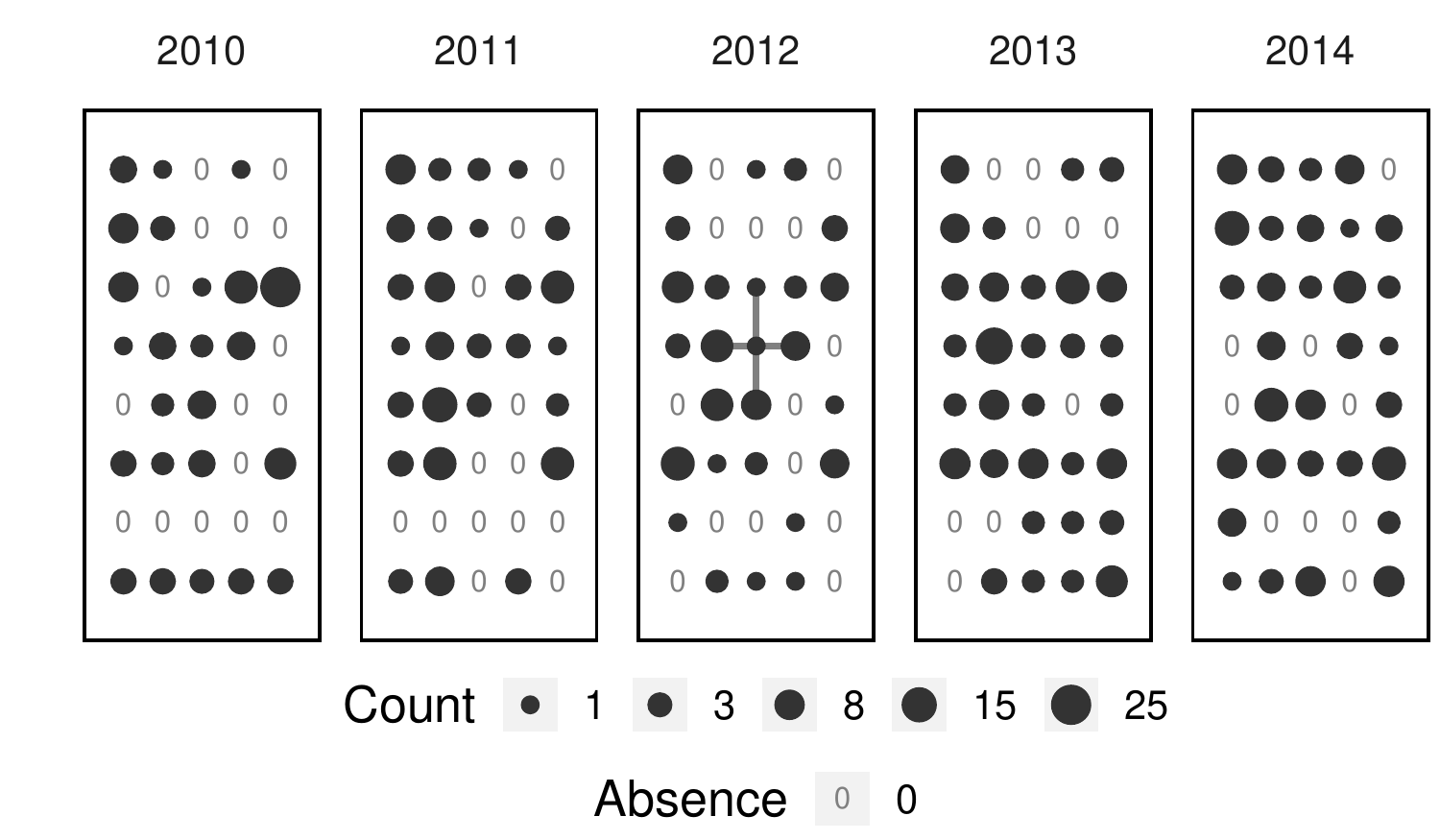}
    \caption{Graphic representation of the spatial distribution of counts of {\it Nenia tridens} and on the sampling lattice for each of 5 illustrative years. Circles represent counts; Grey line represents the neighborhood structure adopted.}
    \label{fig: latice}
\end{figure}

Between $2000$ and $2017$, data on counts (minimum known alive) of \textit{Nenia tridens} were quantified on the LFDP were obtained for the tabonuco forest, see Figure~\ref{fig: latice}. The LFDP is a 16ha rectilinear grid that comprises an 8 x 5 lattice of 40 points (circles of 3m radius), with 60m spacing between adjacent points \citep{willig1998long}. A suite of covariates characterized each of the 40 points and represent habitat characteristics. Some varied in space but not time: Elevation (meters above sea level) and slope (inclination of land in degrees). As a consequence of disturbance and succession, others varied in space and time: density of vegetation (foliar intercepts by plant species, regardless of species identity, in the understory), density of Sierra Palm (foliar intercepts by {\it Prestoea acuminata} in the understory), litter cover (ordinal representation of amount of litter on the forest floor, from 0-2), and canopy openness (estimate of penetration of light to forest understory). To avoid computational problems, all covariates were centered and scaled so that interpretations involve deviations from the mean. In Section~\ref{sec:application} we use this dataset to illustrate our methodology and the dependence structure in the space-time context.


\section{Dependence structure} \label{sec:dependence_structure}

Random variables observed in different regions of space are common in a variety of disciplines and there are many ways to model the spatial dependence among these observations. For areal data, when observations represent a well-defined region, the most traditional model used to capture spatial dependence is the CAR model. In a multivariate context, there is no such agreement in a ``traditional'' model but the MCAR model represents a natural extension of the CAR model. An advantage of MCAR models is to have non-separable dependence structures being a more realistic way to model the relationship between multiple outcomes in a spatial domain in comparison to separable models \citep{rodrigues2010class}. However, its covariance structure is not simple and does not provide an intuitive interpretation regarding the conditional mean and variance of the prior model.    
\subsection{Conditional Autoregressive} \label{subsection: dependence_structure_CAR}

The CAR model is commonly used to model areal data, with each region is represented by one observation. Let $Y_1, \ldots, Y_n$ be a random variable in $n$ regions and let $\mathbf{\Theta} = \PC{\theta_1, \ldots, \theta_n}'$ be random effects with zero mean related to the $n$ regions. The CAR model is specified by the following conditional distributions
\begin{equation*} 
 \PC{\theta_i|\mathbf{\theta_{-i}}} \sim N\PC{\sum_{j \sim i}b_{ij}\theta_j, \tau_i^2},
\end{equation*}
where $\mathbf{\theta_{-i}}$ represents the vector $\mathbf{\Theta}$ without the $i$-th element, $b_{ij}$ is a weight that relates the random effects of regions $i$ and $j$, and $j \sim i$ indicates that region $j$ is a neighbor of region $i$. By Brook's Lemma \citep{brook1964distinction} one can show that the joint distribution of $\mathbf{\Theta} = \PC{\theta_1, \ldots, \theta_n}'$ is 
\begin{equation} \label{eq: CAR_improprio}
 \pi\PC{\bm{\Theta}} \propto \exp\chav{-\frac{1}{2}\bm{\Theta}'\mathbf{D}^{-1}\PC{\mathbf{I_n}-\mathbf{B}}\bm{\Theta}},
\end{equation}
where $\mathbf{D}$ is a diagonal matrix with entries $\tau_i^2$, $i = 1, \ldots, n$ and $\mathbf{B}$ is a $n\times n$ matrix with values $b_{ij}$.

Equation~\eqref{eq: CAR_improprio} resembles the kernel of a multivariate Gaussian distribution with mean $\mathbf{0}$  and covariance structure $\bm{\Sigma} = \PC{\mathbf{I_n}-\mathbf{B}}^{-1}\mathbf{D}$. Nevertheless, it is necessary to guarantee symmetry and the positive definiteness of $\bm{\Sigma}$ for Equation~\eqref{eq: CAR_improprio} to represent a valid Gaussian distribution.

To guarantee symmetry, $\mathbf{B}$ must be such that $\frac{b_{ij}}{\tau_i} = \frac{b_{ji}}{\tau_j}$ $\forall i, j$. A common way to guarantee symmetry is by defining an adjacency matrix $\mathbf{W}$ where $w_{ij} \neq 0$ if and only if $j \sim i$. Thus, $\mathbf{W}$ comprises zeros and ones, where $1$ indicates that areas $i$ and $j$ are neighbors, and $0$ represent regions that are not considered neighbors. Next, define $b_{ij} = \frac{w_{ij} }{w_{i+}}$, where $w_{i+}$ is the sum of the elements of the $i$th row of matrix $\mathbf{W}$, that means, $w_{i+}$ is the number of neighbors of region $i$. And, finally, define that the marginal variance at each region is given by $\tau_i^2 = \frac{\tau^2}{w_{i+}}$. 

In this formulation, $\bm{\Sigma}^{-1} = \frac{1}{\tau^2}(\mathbf{D_w}-\mathbf{W})$, where $\mathbf{D_w}$ is a diagonal matrix with values $w_{i+}$. This assumption only guarantees that $\bm{\Sigma}^{-1}$ is symmetric, but, it is not necessarily positive definite, and so, have no proper joint distribution. This formulation of the CAR model is known as Intrinsic Conditional Autoregressive model (ICAR) and will be denoted as CAR($1$, $\tau^2$).

One can make the previous formulation proper by representing $\bm{\Sigma}^{-1} = \frac{1}{\tau^2}(\mathbf{D_w}-\rho \mathbf{W})$, where $\rho \in (\lambda^{-1}_{min}$, $\lambda^{-1}_{max})$ and $\lambda_{min}$ and $\lambda_{max}$ are the smallest and largest eigenvalues of $\mathbf{D^{\frac{1}{2}}_w}\mathbf{W}\mathbf{D^{\frac{1}{2}}_w}$, respectively \citep{banerjee2014hierarchical}. This configuration will be denoted as CAR($\rho$, $\tau^2$), where $\rho$ is a spatial dependence parameter.

\subsection{Multivariate Conditional Autoregressive} \label{subsection:dependence_structure_MCAR}

A straightforward extension of the CAR model occurs when more than one dependent variable is observed over the same region. This family of multivariate models is known as Multivariate Conditional Autoregressive models \citep[MCAR,][]{gelfand2003proper}.  

Let $n$ be the number of regions of interest and $p$ the number of variables observed. Define $\mathbf{Y_1}$ and $\mathbf{\Theta_1}$ a vector of observations and spatial random effects respectively ordered by region, then
\begin{center}
$\mathbf{Y_1}$ = $\PC{Y_{11}, Y_{12}, \ldots, Y_{1p}, Y_{21}, \ldots, Y_{2p}, \ldots, Y_{n1} \ldots, Y_{np}}$,
$\mathbf{\Theta_1} = \PC{\theta_{11}, \theta_{12}, \ldots, \theta_{1p}, \theta_{21}, \ldots, \theta_{2p}, \ldots \theta_{np}}$.
\end{center}
Now define $\mathbf{Y_2}$ an observation vector and $\mathbf{\Theta_2}$ a spatial random effect sorted by variable, thus
\begin{center}
$\mathbf{Y_2}$ = $\PC{Y_{11}, Y_{21}, \ldots, Y_{n1}, Y_{12}, \ldots, Y_{n2}, \ldots, Y_{1p} \ldots, Y_{np}}$,

$\mathbf{\Theta_2} = \PC{\theta_{11}, \theta_{21}, \ldots, \theta_{n1}, \theta_{12}, \ldots, \theta_{n2}, \ldots \theta_{np}}$.
\end{center}

Generally, the MCAR model can be defined by conditional distributions for $\mathbf{\theta_{ij}}$ as
\begin{equation*} 
\PC{\mathbf{\theta_{ij}}|\mathbf{\theta_{-ij}}} \sim N_{np}\PC{\sum_{k,l \sim i,j}b_{ij,kl}\mathbf{\theta_j}, \tau_{ij}},
\end{equation*}
where $ij \sim kl$ are defined as the neighbors of a variable $j$ in region $i$ with a variable $l$ in region $k$. Applying Brook's Lemma, it is possible to calculate the joint distribution of $\mathbf{\theta}$ as 
\begin{equation} \label{eq:MCAR_Brook}
\Pi\PC{\mathbf{\Theta}} \propto \exp\chav{-\frac{1}{2}\mathbf{\Theta}'\mathbf{Q}\mathbf{\Theta}},
\end{equation}
with $q_{ij,kl} = \frac{-b_{ij,kl}}{\tau_{ij}}$. Therefore, as in the univariate case, it is necessary to guarantee that $\mathbf{Q}$ is symmetric and positive definite. Different choices of the coefficients $b_{ij,kl}$ and $\tau_{ij}$ determine the methodologies that are available from the literature. 

Given the general representation of Equation~\eqref{eq:MCAR_Brook}, an alternative way of interpreting and understanding the multivariate distribution is considering its conditional mean 
\begin{equation} \label{eq:sain_esperanca}
E\PC{\theta_{ij}|\theta_{-\{ij\}}} = \underbrace{\sum_{k \neq i}b_{ij,kj}\theta_{kj}}_{\mathbb{A}} + \underbrace{\sum_{l \neq j}b_{ij,il}\theta_{il}}_\mathbb{B} + \underbrace{\sum_{k,l \neq i,j}b_{ij,kl}\theta_{kl}}_\mathbb{C},
\end{equation}
and conditional variance
\begin{equation} \label{eq:sain_variancia}
Var\PC{\theta_{ij}|\theta_{-\{ij\}}} = \tau^2_{ij},
\end{equation}
where $b_{ij,kl}$ is the associated weights of regions $i$ and $k$ according to variable $j$ and $l$. This representation allows for a direct interpretation of the sums $\mathbb{A}$, $\mathbb{B}$, and $\mathbb{C}$ in Equation~\eqref{eq:sain_esperanca}: $\mathbb{A}$ measures the dependence between variables in the same region, $\mathbb{B}$ measures spatial dependence within the same variable, and $\mathbb{C}$ measures the spatial dependence between different variables. 

Next, we revisit many of the available MCAR proposals in the literature to show its representation in the general formulation of Equation~\eqref{eq:MCAR_Brook} and to provide a more intuitive interpretation using the conditional mean structure in Equation~\eqref{eq:sain_esperanca}. 
\begin{itemize}
\item {\bf \cite{gelfand2003proper} and \cite{carlin2003hierarchical}}:
\end{itemize}
Ordering the data by region the authors define the matrix $\mathbf{Q_1}$ as:
\begin{equation} \label{eq: MCAR_gelfand_carlin_regiao}
\mathbf{Q_1} = \PC{\mathbf{D_w} - \rho \mathbf{W}}\otimes\bm{\Lambda},
\end{equation}
where, $\bm{\Lambda}$ is a $p\times p$ matrix. Given their proposal, $b_{ij,kl}$ and $\tau_{ij}$ are given by:
$$b_{ij,kl} = \left\{\begin{array}{lc}
\rho\frac{w_{ik}}{w_{i+}}, & \mbox{se} \quad j = l \quad \mbox{e} \quad i \neq k,\\
\frac{-\Lambda_{jl}}{\Lambda_{jj}}, & \mbox{se} \quad j \neq l \quad \mbox{e} \quad i = k,\\
\rho\frac{w_{ik}}{w_{i+}}\frac{\Lambda_{jl}}{\Lambda_{jj}}, & \mbox{se} \quad j \neq l \quad \mbox{e} \quad i \neq k.
\end{array}\right. \mbox{And}
$$
$$\tau_{ij} = \frac{1}{w_{i+}\Lambda_{jj}}.$$
When the data are ordered by variable, we have $\mathbf{Q_2} = \bm{\Lambda} \otimes \PC{\mathbf{D_w} - \mathbf{\rho} \mathbf{W}},$ with coefficients $b_{ij,kl}$ and $\tau_{ij}$ identical to representations in the previous case.

After determining the weighting coefficient $b_{ij,kl}$, we can use Equation~\eqref{eq:sain_esperanca} to see how the proposed model affects the conditional mean via
$$E\PC{\theta_{ij}|\theta_{-\{ij\}}} = \sum_{k \neq i}\rho\frac{w_{ik}}{w_{i+}}\theta_{kj}-\sum_{l \neq j}\frac{\Lambda_{jl}}{\Lambda_{jj}}\theta_{il}+\sum_{k,l \neq i,j}\rho\frac{w_{ik}}{w_{i+}}\frac{\Lambda_{jl}}{\Lambda_{jj}}\theta_{kl}.$$
In the first summation, $\rho$ is the dependence parameter as in the univariate case, but it also appears in the third summation, as a smoothing parameter on the cross-variable dependence. Thus, $\rho$ is now directly related to two quantities making it challenging to interpret its contribution to the model. The second summation represents a smoothing in the dependence between variables, its negative sign makes interpretation unclear.
\begin{itemize}
   \item {\bf \cite{jin2007order}}:
\end{itemize}
The representation of \cite{jin2007order} has a restriction in order and can be sorted only by variable, thus the authors define the matrix $\mathbf{Q_2}$ as
\begin{equation} \label{eq: MCAR_Brook}
\mathbf{Q_2} = \left(
  \begin{array}{ccccccccc}
   (D_w - \rho_{11}W)\Lambda_{11} & \ldots & (D_w - \rho_{1p}W)\Lambda_{1p} \\
   \vdots                         & \ddots & \vdots                         \\
   (D_w - \rho_{1p}W)\Lambda_{1p} & \ldots & (D_w - \rho_{1p}W)\Lambda_{pp} \\   
  \end{array}
\right).
\end{equation}
With this structure, we can find $b_{ij,kl}$ and $\tau_{ij}$ as
$$b_{ij,kl} = \left\{\begin{array}{lc}
\rho_{jj}\frac{w_{ik}}{w_{i+}}, & \mbox{se} \quad j = l \quad \mbox{e} \quad i \neq k,\\
-\frac{\Lambda_{jl}}{\Lambda_{jj}}, & \mbox{se} \quad j \neq l \quad \mbox{e} \quad i = k,\\
\rho_{jl}\frac{w_{ik}}{w_{i+}}\frac{\Lambda_{jl}}{\Lambda_{jj}}, & \mbox{se} \quad j \neq l \quad \mbox{e} \quad i \neq k,
\end{array}\right.
$$
$$\tau_{ij} = \frac{1}{w_{i+}\Lambda_{jj}}.$$
Like the previous formulation, this representation does not provide a straightforward interpretation of the parameter. From the conditional mean we have
\begin{equation}\label{eq:Gelfand2003}
E\PC{\theta_{ij}|\theta_{-\{ij\}}} = \sum_{k \neq i}\rho_{jj}\frac{w_{ik}}{w_{i+}}\theta_{kj}-\sum_{l \neq j}\frac{\Lambda_{jl}}{\Lambda_{jj}}\theta_{il}+\sum_{k,l \neq i,j}\rho_{jl}\frac{w_{ik}}{w_{i+}}\frac{\Lambda_{jl}}{\Lambda_{jj}}\theta_{kl},
\end{equation}
which is similar to the parametrization presented by \cite{gelfand2003proper} and \cite{carlin2003hierarchical}, but is a little more flexible since it allows the application of different weights $\rho_{jl}$ instead of only a unique $\rho$. However, as presented in Equation~\eqref{eq:Gelfand2003}, it still has a negative sign in the second summation, making interpretation a challenge.
\begin{itemize}
   \item {\bf \cite{sain2011spatial}}:
\end{itemize}

The authors define $\mathbf{Q_1}$ as
\begin{equation} \label{eq: MCAR_sain}
\mathbf{Q_1} = \PC{I_n \otimes \tau^{-\frac{1}{2}}}\PC{I_n \otimes A - W \otimes B}\PC{I_n \otimes \tau^{-\frac{1}{2}}},
\end{equation}
where
\small
\begin{multicols}{3}
$$\mathbf{\Lambda} = \left(
  \begin{array}{ccc}
    \Lambda^2_1 & \ldots & 0      \\
    \vdots & \ddots & \vdots \\
    0      & \ldots & \Lambda^2_p \\
  \end{array}
\right),$$
$$\mathbf{A} = \left(
  \begin{array}{ccc}
    1           & \ldots & -\rho_{1p} \\
    \vdots      & \ddots & \vdots    \\
    -\rho_{1p}  & \ldots & 1         \\
  \end{array}
\right),$$
$$\textbf{B} = \left(
  \begin{array}{ccc}
    \rho_{1}   & \ldots &  \psi_{1p}  \\
    \vdots       & \ddots & \vdots   \\
    \psi_{1p}   & \ldots &  \rho_{p}  \\
  \end{array}
\right).$$
\end{multicols}
\normalsize
From this representation $b_{ij,kl}$ and $\tau_{ij}$ are defined by
$$b_{ij,kl} = \left\{\begin{array}{lc}
\rho_{j}w_{ik}, & \mbox{se} \quad j = l \quad \mbox{e} \quad i \neq k,\\
\rho_{jl}\frac{\Lambda_{j}}{\Lambda_{l}}, & \mbox{se} \quad j \neq l \quad \mbox{e} \quad i = k,\\
w_{ik}\psi_{jl}\frac{\Lambda_{j}}{\Lambda_{l}}, & \mbox{se} \quad j \neq l \quad \mbox{e} \quad i \neq k,
\end{array}\right.
$$
and $\tau_{ij} = \Lambda^2_j$. The equation for the conditional mean is now 
\begin{equation}\label{eq:Sain2011}
E\PC{\theta_{ij}|\theta_{-\{ij\}}} = \sum_{k \neq i}\rho_{j}w_{ik}\theta_{kj}+\sum_{l \neq j}\rho_{jl}\frac{\Lambda_{j}}{\Lambda_{l}}\theta_{il}+\sum_{k,l \neq i,j}\psi_{jl}w_{ik}\frac{\Lambda_{j}}{\Lambda_{l}}\theta_{kl}.
\end{equation}
This representation has four salient differences compared to the previous two approaches: (1) the contribution of the second summation is positive; (2) each variable is characterized by a different spatial parameters ($\rho_j$) in the first summation; (3) different parameters ($\rho_{jl}$) accommodate the dependence between variables in the second summation; and (4) a smoothing parameter ($\psi_{jl}$) in the third summation controls cross-dependence between variables in the $i$th region and its spatial neighbors.

Although the model seems flexible, its interpretation is not trivial because the conditional mean depends on the $\Lambda_{i}$ parameters. Moreover, the summations are not weighted, which implies that if the number of neighbors of one area increases, the expected mean will always increase, making this assumption unrealistic. Another drawback of this dependence structure is that is not easy to guarantee that the proposed matrix is positive definite.

\subsubsection{A New Alternative}
\label{s:apw}
With the last parametrization in mind, we present a new formulation that maintains flexibility and, motivated by the conditional mean and variance representation, allows for a direct interpretation of parameters in the model and provides a direct theoretical space for the parameters that guarantee the positive definite nature of the dependence matrix. 

\begin{figure}
\center
\subfloat[]{
\includegraphics[width = 0.3\textwidth]{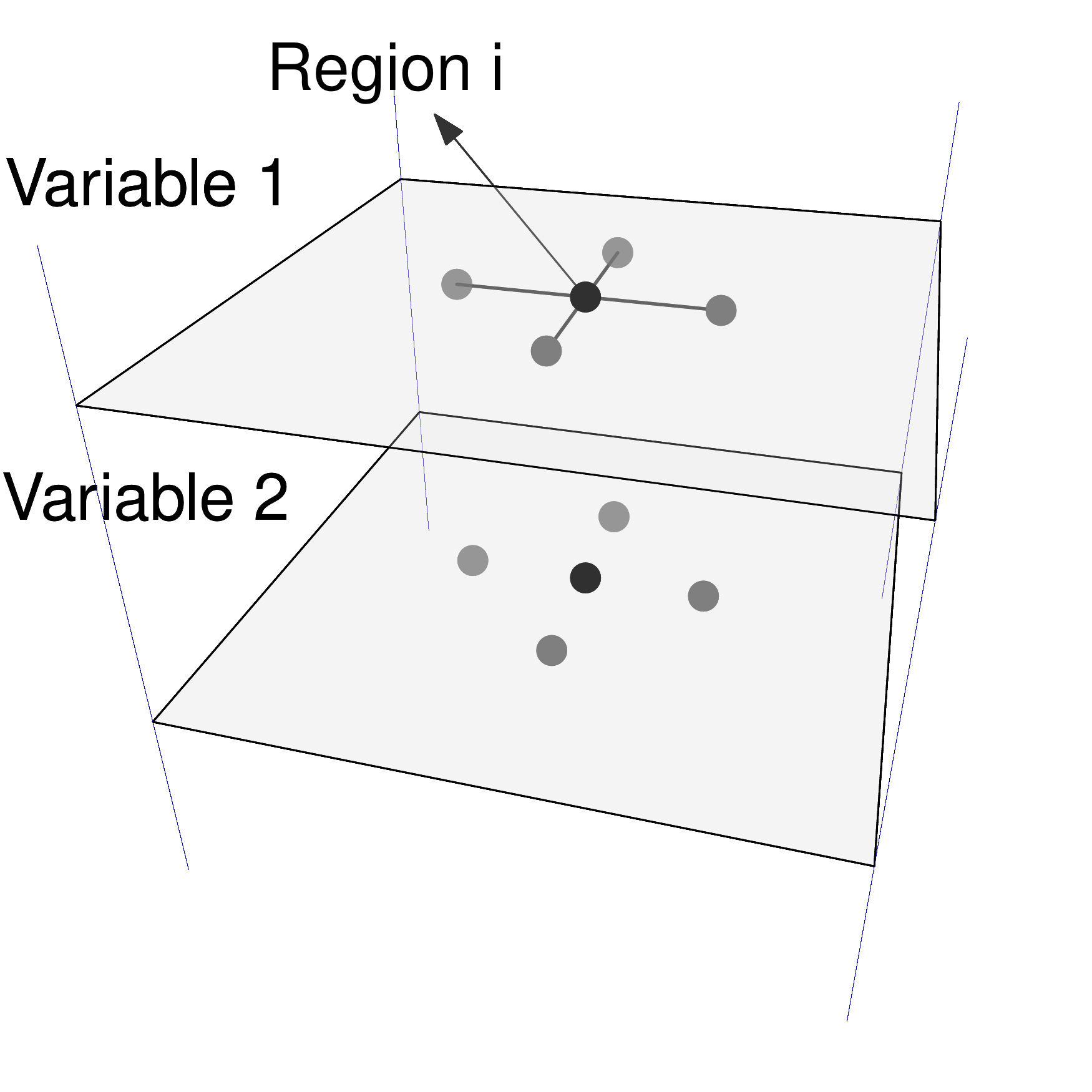}
\label{fig:tipos_vizinhanca_espacial}
}
\subfloat[]{
\includegraphics[width = 0.3\textwidth]{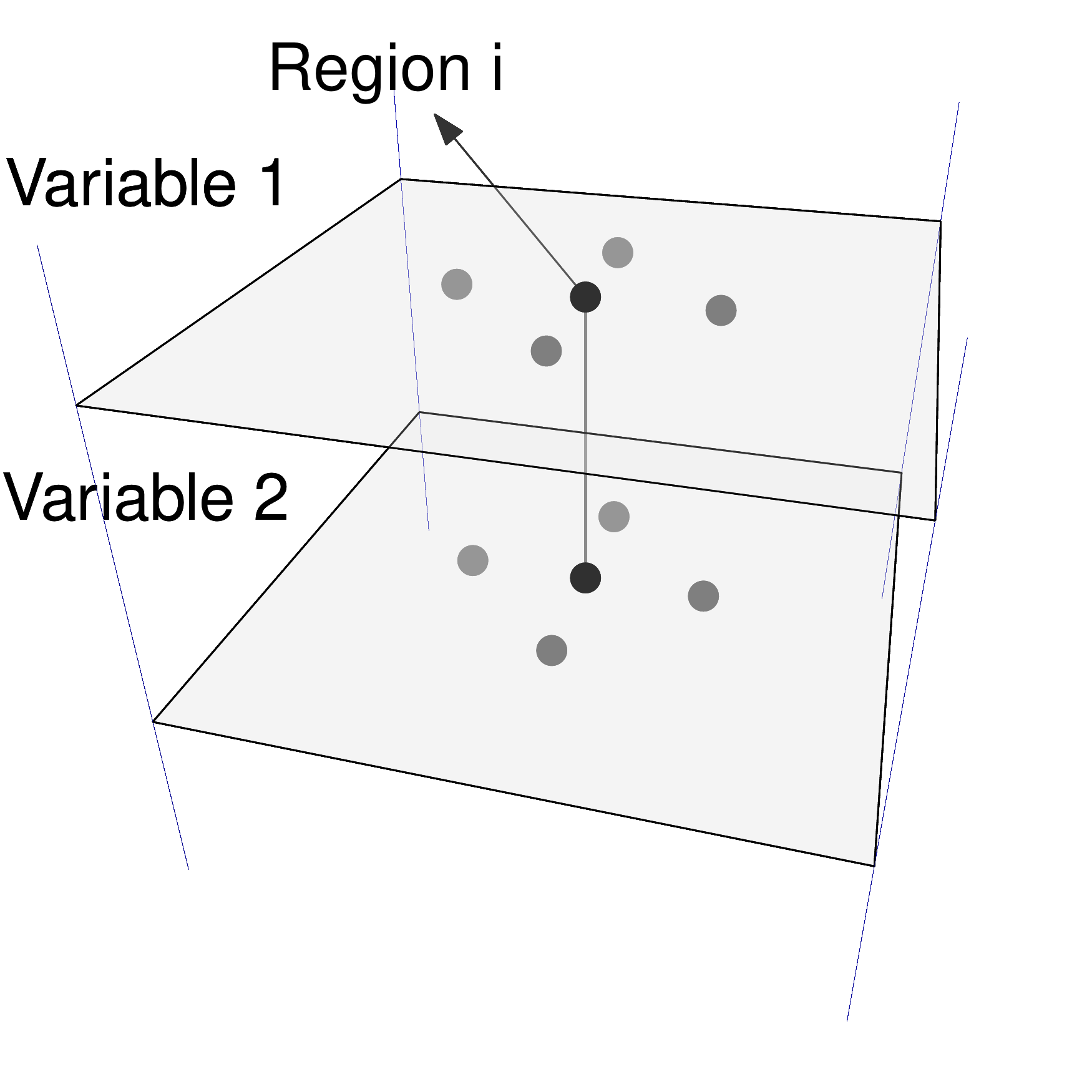}
\label{fig:tipos_vizinhanca_temporal}
}
\subfloat[]{
\includegraphics[width = 0.3\textwidth]{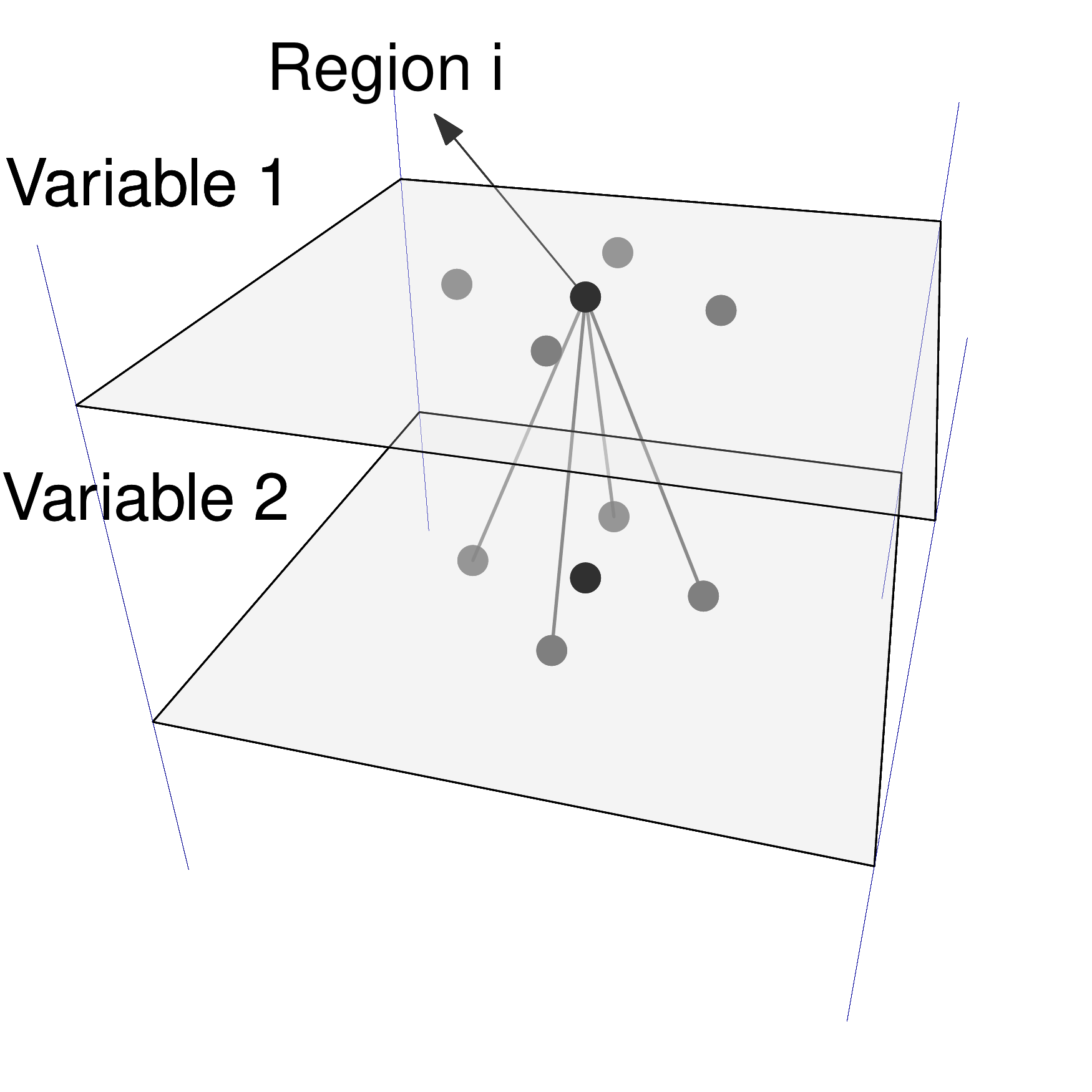}
\label{fig:tipos_vizinhanca_espaco_temporal}
}
\caption{Types of neighborhood. (a): spatial neighborhood within the variable. (b): neighborhood between the same region and different variables. (c): neighborhood between a particular region and its spatial neighbor regions across different variables.} 
\label{fig:tipos_vizinhanca}
\end{figure}

Three types of neighborhoods should be considered in characterizing the dependence structure of a model: (1) spatial neighbors of region $i$ (Figure~\ref{fig:tipos_vizinhanca_espacial}), (2) neighbors of the same region $i$ between variables (Figure~\ref{fig:tipos_vizinhanca_temporal}), (3) spatial neighbors of region $i$ across different variables (Figure~\ref{fig:tipos_vizinhanca_espaco_temporal}).

After defining these neighborhood structures and following Equations~\eqref{eq:sain_esperanca} and \eqref{eq:sain_variancia}, we now define the conditional mean and variance as  
\begin{equation*}
E\PC{\theta_{ij}|\theta_{-\{ij\}}} = \sum_{k \neq i}\rho^{s}_{j} \frac{w_{ik}}{D_{ij}}\theta_{kj}+\sum_{l \neq j}\rho^{p}_{jl} \frac{v_{jl}}{D_{ij}}\theta_{il}+\sum_{k,l \neq i,j}\rho^{sp}_{ij,kl}\frac{w_{ik}v_{jl}}{D_{ij}}\theta_{kl},
\end{equation*}
and
$$Var\PC{\theta_{ij}|\theta_{-\chav{ij}}} = \frac{\tau}{D_{ij}},$$
where $D_{ij}$ represents the total number of neighbors of region $i$ with regard to variable $j$. This new constant is a weighting term; the correction is necessary so that the conditional mean can be interpreted as an average term that does not always increase with the number of neighbors (in contrast with Equation~\eqref{eq:Sain2011}). It is also essential to easily define the parametric space for the matrix parameters to ensure its positive definite nature as will be explained later. 

The proposed model is intuitive and provides direct interpretation for all parameters. Summation $\mathbb{A}$ in Equation~\eqref{eq:sain_esperanca} controls for the spatial dependence in the variable, thus, $\rho^{s}_{j}$ measure the spatial dependence within variable $j$ (Figure~\ref{fig:tipos_vizinhanca_espacial}), $\rho^p_{jl}$ capture the association between the $j$th and $l$th variable at the same location (Summation $\mathbb{B}$ in Equation~\eqref{eq:sain_esperanca} represented in Figure~\ref{fig:tipos_vizinhanca_temporal}), and, in Summation $\mathbb{C}$ in Equation~\eqref{eq:sain_esperanca}, $\rho^{sp}_{ij,kl}$ controls for the dependence between the $j$ variable in region $i$ with its $k$ spatial neighbors considering other variables $l\neq j$ (Figure~\ref{fig:tipos_vizinhanca_espaco_temporal}).

From this representation, we calculate the joint model. 
\begin{proposition} \label{prop:model}
Let $\bm{\rho}^s = \mbox{diag}(\rho^s_1, \ldots, \rho^s_p)$, $V_p$ a $p \times p$ matrix with entries $\rho^p_{ij}$ and $\rho^p_{ii} = 0$, and $\bm{U}_{sp}$ a $np \times np$ matrix with entries $\rho^{sp}_{ij,kl}$. With this specification, the $\bm{Q}$ structure is given by:
\begin{equation} \label{eq:MCAR_douglas_marcos1}
\bm{Q}_1 = \frac{1}{\tau}\PC{D - W\otimes \mathbf{\rho^s} - I_{n}\otimes V_p - U_{sp}\odot (W\otimes V)},
\end{equation}
where $\odot$ is the Hadamard product between two matrices. In this case, we get $b_{ij,kl}$ and $\tau_{ij}$ as:
$$b_{ij,kl} = \left\{\begin{array}{lc}
\rho^s_j\frac{w_{ik}}{D_{ij}}, & \mbox{se} \quad j = l \quad \mbox{e} \quad i \neq k,\\
\rho^p_{jl}\frac{v_{jl}}{D_{ij}}, & \mbox{se} \quad j \neq l \quad \mbox{e} \quad i = k,\\
\rho^{sp}_{ij, kl}\frac{w_{ik}v_{jl}}{D_{ij}}, & \mbox{se} \quad j \neq l \quad \mbox{e} \quad i \neq k,
\end{array}\right.
$$
$$\tau_{ij} = \frac{\tau}{D_{ij}}.$$
And if ordered by variable instead of by region, we have
\begin{equation} \label{eq: MCAR_douglas_marcos1}
\bm{Q}_2 = \frac{1}{\tau}\PC{D - V_p\otimes I_{n} - \mathbf{\rho^s} \otimes W - U_{sp}\odot (V\otimes W)}.
\end{equation}
\end{proposition}
Proposition~\ref{prop:model} provides a valid multivariate distribution, with $\bm{Q}$ being symmetric and positive definite. The positive definiteness of $\bm{Q}$ is guaranteed by the diagonal dominance criterion with all its diagonals elements positive. A square matrix is diagonally dominant if for every row the value of the diagonal element is larger than the summation of the absolute values out of the diagonal:
$$q_{ii} > \sum_{j \neq i} |q_{ij}| ~~ \forall i,$$
where $q_{ij}$ denotes the entry at the $i$-th line and $j$-th column. The diagonal dominance is a sufficient but not necessary condition for a symmetric matrix to be positive definite. 
Therefore, for this definition, the parametric space of $\rho$'s directly depends on the neighborhood structure since for $\bm{Q}$ the $q_{ij}$'s are functions of the $\rho$'s parameters. 


After defining the dependence structures for the multivariate case, it is clear that the spatio-temporal setup can be seen as an equivalent case where instead of having $p$ variables in a map, we have one variable observed over the whole map in a discrete period of $T$ times.  

\section{Spatio-temporal modeling using TGMRFs} \label{sec:TGMRF}

In the spatial setup, the Transformed Gaussian Markov Random Field (TGMRF) was proposed as a flexible alternative to GMRF \citep{prates2015transformed}. In this class, the marginal distribution is chosen accordingly to each application, providing flexibility in being capable of accommodating asymmetry, heavy tails or other characteristics, thereby maintaining many desirable properties of the GMRF \citep{prates2011link}. In summary, the TGMRF uses a copula approach to separate the marginal structure of the model from the dependent one. This is an interesting component because by construction \citep{hughes2015copcar, prates2015transformed}, it does not suffer from spatial confounding \citep[][and others]{reich2006effects, hodges2010adding, hanks2015restricted, prates2018alleviating} and as a direct consequence, it will not suffer spatio-temporal confounding between fixed and random effects.   

In more detail, a TGMRF is obtained by transforming the marginal distribution of the GMRFs to a desired one. Let $\bm{\epsilon} = \PC{\epsilon_1, \ldots, \epsilon_n}'$ be a multivariate normal vector with mean $\bm{0}$ and sparse correlation matrix $\bm{\varXi}$, $\bm{\epsilon} \sim N_n(\bm{0}, \bm{\varXi})$, consequently $\bm{\epsilon}$ is a GMRF. Let 
$\bm{Z} = \PC{Z_1, \ldots, Z_n}'$ and $Z_i = F_i^{-1}\chav{\Phi\PC{\epsilon_i}}$, $i = 1, \ldots, n$, where $F_i(x)$ is the cumulative distribution function (cdf) of an absolutely continuous function in respect to the support of $x$ and $\Phi$ is the cdf of the $N\PC{0, 1}$. 
So, each $Z_i$ has marginal distribution $f_i$ (probability density distribution (pdf) of $F_i$) and jointly a TGMRF with marginals $\bm{F}$ and dependence structure $\bm{\varXi}$, denoted by $\bm{Z} \sim TGMRF_n(\bm{F}, \bm{\varXi})$. The $\bm{Q} = \bm{\varXi}^{-1}$ brings a more intuitive interpretation of the conditionals distribution of $\bm{Z}$, and we parametrize the TGMRF by its precision matrix $\bm{Q}$ and denoted by $\bm{Z} \sim TGMRF_n(\bm{F}, \bm{Q})$.

TGMRFs can be used to directly model Poisson intensities or Bernoulli rates, taking into account a marginal distribution of interest and spatial dependence \citep{prates2015transformed}. For example, in a Poisson regression, the TGMRF is defined as a joint distribution for $\bm{\mu}$ as
\begin{equation}\label{eq: TGRF}
 \bm{\mu} \sim TGMRF_n(\mathbf{F}, \bm{Q}),
\end{equation}
where $\bm{F} = \PC{F_1, \ldots, F_n}$, $F_i$ is a desired and adequate cdf for the marginal distribution of $\mu_i$ with pdf $f_i$ and dependence matrix $\bm{Q}$.

From a spatio-temporal perspective let $\bm{Y} = \PC{\bm{Y}_{1}', \bm{Y}_{2}', \ldots, \bm{Y}_{T}'}'$ be a random vector observed at $T$ times and $n$ regions with $\bm{Y}_{t} = \PC{Y_{1t}, Y_{2t}, \ldots, Y_{nt}}$ for $t = 1, \ldots, T$. The $nT \times q$ covariate matrix is defined as $\bm{X} = \PC{\bm{X}_{1}, \bm{X}_{2}, \ldots, \bm{X}_{q}}$ with $\bm{X}_{j} = \PC{X_{11}, \ldots, X_{n1}, X_{12}, \ldots, X_{nT}}$ for $j = 1, \ldots q$ and random effects $\bm{\epsilon} = \PC{\bm{\epsilon_{1}}', \bm{\epsilon}_{2}', \ldots, \bm{\epsilon}_{T}'}'$ with $\bm{\epsilon}_{t} = \PC{\epsilon_{1t}, \epsilon_{2t}, \ldots, \epsilon_{nt}}$ following a $N_{nt}(\bm{0}, \bm{Q})$.

If the distribution of the random variables $Y_{it}$ belongs to the exponential family with mean $\mu_{it} = E(Y_{it}|\bm{X},\epsilon_{it})$, then the joint distribution of $\bm{\mu}$ can be modeled by a TGMRF as 
\begin{equation*}
\bm{\mu} \sim TGMRF_{nt}(\mathbf{F}, \bm{Q}),
\end{equation*}
where $\bm{F} = \PC{F_{11}, \ldots, F_{1T}, F_{21}, \ldots, F_{nT}}$, $F_{it}$ is the cdf related to the marginal distribution of $\mu_{it}$ and $\bm{Q}$ is the dependence matrix of $\bm{\mu}$.

Let $\bm{\xi} = \PC{\bm{\beta},\bm{\rho}, \bm{\nu}}$, where $\bm{\rho} = \PC{\rho_s, \rho_t, \rho_{st}}$ and $\bm{\nu}$ are hyperparameters of the distribution $\bm{F}$. A spatio-temporal hierarchical TGMRF model can be defined as:
\begin{eqnarray} \nonumber
\displaystyle Y_{it}|\mu_{it} &\sim& \pi\PC{y|\mu_{it}}, i = 1, \ldots, n; t = 1, \ldots T,\\
\displaystyle \bm{\mu} &\sim& TGMRF_{nt}\PC{\bm{F}_{\bm{\xi}, \bm{X}},\bm{Q}_{\bm{\rho}}} \label{eq:STTGMRF}\\ \nonumber
&&\displaystyle \pi\PC{\bm{\beta}}, \displaystyle \pi\PC{\bm{\nu}}, \pi\PC{\bm{\rho}},
\end{eqnarray}
where $\bm{F}_{\bm{\xi}, \bm{X}}$ may depend on the covariates $\bm{X}$, regression coefficient vector $\bm{\beta}$, dispersion parameter(s) $\bm{\nu}$ and spatial, temporal and spatio-temporal parameters $\rho_s$, $\rho_t$ and $\rho_{st}$, respectively. The precision matrix $\bm{Q}_{\bm{\rho}}$ will depend on only the dependence parameters $\bm{\rho}$.

As for the spatial setting, this formulation will not suffer from spatio-temporal confounding because it separates the marginal effects of the dependence structure. Moreover, it allows for flexible representations of marginals distributions. To avoid over parametrization and to construct a dependence matrix capable of carrying the flexibility of model \eqref{eq:STTGMRF} combined with an intuitive parameter interpretation, we propose to use the dependence matrix $\bm{Q}$ in Section~\ref{s:apw} with $\rho^{s}_{j} \equiv \rho_s$, $\rho^p_{jl} \equiv \rho_t$ and $\rho^{sp}_{ij,kl} \equiv \rho_{st}$. With this formulation, we have the conditional mean and variance defined as  
$$E\PC{\theta_{ij}|\theta_{-\{ij\}}} = \sum_{k \neq i}\rho_{s} \frac{w_{ik}}{D_{ij}}\theta_{kj}+\sum_{l \neq j}\rho_{t} \frac{v_{jl}}{D_{ij}}\theta_{il}+\sum_{k,l \neq i,j}\rho^{st}\frac{w_{ik}v_{jl}}{D_{ij}}\theta_{kl},$$
and
$$Var\PC{\theta_{ij}|\theta_{-\chav{ij}}} = \frac{\tau}{D_{ij}},$$
where $\rho_s$ accommodate the spatial dependence between regions, $\rho_t$ represent the temporal dependence between time $t$ and its previous ($t-1$) and its next ($t+1$), mimicking an autoregressive model in time with order $1$ and
$\rho_{st}$ model the dependence between area $i$ in time $t$ and its spatial neighbors in time $t-1$ and $t+1$.

\subsection{Marginal models and inference} \label{subsec:marginais_inferencia}

When a traditional GLMM is used to fit a Poisson model, it is common to use the log-link function. It is easy to prove that under this link function the marginal distribution for the conditional mean is log-normal. 

Under TGMRFs models, we can set the family, mean and variance of these distributions. Table~\ref{tab: models} shows the means and variances of marginal distributions used in this work. 
\begin{table}
\centering
\caption{Mean and variance of marginal models.}\label{tab: models}
\resizebox{.99\textwidth}{!}{
\begin{tabular}{cccc}
  \hline
  Model & Model name & $E\PC{\mu_{ij}}$ & V$\PC{\mu_{ij}}$ \\
  \hline
  Gamma Independent & GI & $\exp\chav{\bm{X_{ij}\bm{\beta}}}$ & $\frac{1}{\nu}$ \\
  Gamma Scale & GSC & $\exp\chav{\bm{X_{ij}\bm{\beta}}}$ & $\frac{1}{\nu}\exp\chav{\bm{X_{ij}\bm{\beta}}}^2$ \\
  Gamma Shape & GSH & $\exp\chav{\bm{X_{ij}\bm{\beta}}}$ & $\frac{1}{\nu}\exp\chav{\bm{X_{ij}\bm{\beta}}}$ \\
  Log-Normal & LN & $\exp\chav{\bm{X_{ij}\bm{\beta}} + 0.5\frac{1}{\nu}\bm{Q^{-1}_{ij, ij}}}$ & $\exp\chav{2\bm{X_{ij}\bm{\beta} + \nu\bm{Q^{-1}_{ij, ij}}}}\PC{\exp\chav{\nu\bm{Q^{-1}_{ij, ij}}}-1}$ \\
  \hline
\end{tabular}}
\end{table}

An equivalent approach to the usual GLMM under log-link function is the Log-Normal model. Other distributions allow flexibility to the model. Importantly, the $\nu$ parameter is not equivalent in these models and we do not expect the same estimation for this parameter under model misspecification.

Let $\mathbf{Y} = \PC{Y_{11}, \ldots, Y_{1t}, Y_{21}, \ldots, Y_{2t}, \ldots, Y_{nt}}$, random variables in $n$ regions and $t$ different times. Then we have $Y_{ij}|\mu_{ij} \sim Poisson(\mu_{ij})$, $\forall i = 1, \ldots, n$ e $j = 1, \ldots, t$. Let $\mathbf{Q_{\bm{\rho}}}$ the structure matrix defined in Equation~\eqref{eq: MCAR_douglas_marcos1} with $\rho^{s}_{j} \equiv \rho_s$, $\rho^p_{jl} \equiv \rho_t$ and $\rho^{sp}_{ij,kl} \equiv \rho_{st}$ and let $\bm{\beta}$ a coefficient vector of dimension $q$. We used a Gibbs Sampling algorithm with Metropolis-Hastings step for each parameter in the modeling. Priors distributions were set to be flat on their domain even for the dependence parameters.

To compare methods we used WAIC \citep{Watanabe}, LPML \citep{Geisser,Dey} and DIC \citep{spiegelhalter2002bayesian}. A broader discussion about the criteria can be found in \cite{gelman2014understanding}.

To allow reproducibility and provide access for a wider range of practitioners, an \texttt{R} package has been created and can be installed following the instructions in the \texttt{TGMRF: Transformed Gaussian Markov Random Fields} repository \url{https://github.com/douglasmesquita/TGMRF}.

\section{Simulation study} \label{sec:simulation}

To evaluate our method, we performed a simulation study. The global sample size is always fixed at $300$ but the spatio-temporal design varies across scenarios. The MCMC setup was calibrated after empirical tests that showed that a chain with $1000$ samples thinned by $10$ to reduce auto-correlation after $5000$ iterations of burn-in is sufficient to achieve convergence and estimate parameters.

As our method is applied for a spatio-temporal setting, we divided our study into three parts. First, we investigated the ability of our method to restore parameters under a situation in which we have temporal but not spatial independence. Second, we explored a scenario where there is spatial but not temporal independence. Finally, we considered a more realistic scenario in which spatial, temporal and spatio-temporal dependence is present.

For all scenarios, data was generated from each one of the models defined in Table~\ref{tab: models}. For each proposed model, $100$ datasets were generated and for all of them, we fitted the dataset using all model proposals. Table \ref{tab: scenarios} shows our parameters in each scenario and the dimension of the lattice used for simulations.
\begin{table}
\caption{Parameters for each simulation scenario and lattice sizes.}\label{tab: scenarios}
\centering
\begin{tabular}{@{}llll@{}}
\toprule
Parameter   & Scenario 1   & Scenario 2   & Scenario 3 \\ \midrule
$\beta_0$   & 1            & 1            & 1          \\
$\beta_1$   & -0.1         & -0.1         & -0.1       \\
$\rho_s$    & 2.18         & 0            & 0.97       \\
$\rho_t$    & 0            & 3.88         & 1.71       \\
$\rho_{st}$ & 0            & 0            & 0.77       \\
\#Rows      & 6            & 6            & 6          \\
\#Columns   & 5            & 5            & 5          \\
\#Times     & 10           & 10           & 10         \\ \bottomrule
\end{tabular}
\end{table}


To demonstrate the accuracy of the method, we present the results of Scenario 3 in Table~\ref{tab: results 3}. Results are illustrated as mode, standard deviations and mean squared errors. 
The different choices for $\nu$ were such that the mean marginal variance $V(\mu_{ij})$ of each model was set around $10$ (as can be seen in  Table~\ref{tab: results 3}).  The point estimates of the parameters are well recovered for the true generating mode with a low MSE (Mean Square Error). Even under model misspecification $\beta_1$ and $\bm{\rho}$ seem to be nicely recovered between the models. Because of the copula separation of the TGMRF the dependence parameters in $\bm{\rho}$ do not depend on the choice of the marginal link. The traditional LN model has not the same marginal mean as the other proposal, for this reason, $\beta_0$ for the LN model is not comparable with the Gamma proposals. The same observation can be made for $\nu$ since they are not comparable along with the models and therefore estimated fairly different between them. 

Similar observations, not shown and available upon request, are made for scenarios 1 and 2. 
Therefore, we conclude that our method can recover spatial, temporal and spatio-temporal characteristics as well as coefficients and scale or variability parameters.
\begin{table}[htb]
\caption{Simulation study for the spatio-temporal scenario. $\nu$ parameter are note comparable across models. Results are shown as Mode (Standard Deviation) and Mean Squared Error (MSE) for 100 simulated datasets.}
\label{tab: results 3}
\resizebox{.99\textwidth}{!}{
\begin{tabular}{@{}lllllllllll@{}}
\toprule
\multirow{3}{*}{\begin{tabular}[c]{@{}l@{}}True \\ model\end{tabular}} & \multirow{3}{*}{Parameters} & \multirow{3}{*}{\begin{tabular}[c]{@{}l@{}}True\\ value\end{tabular}} & \multicolumn{8}{l}{Specified model}                                                           \\ \cmidrule(l){4-11} 
                                                                       &                             &                                                                       & GI           &        & GSC          &        & GSH          &        & LN           &        \\ \cmidrule(l){4-11} 
                                                                       &                             &                                                                       & Mode (SD)    & MSE    & Mode (SD)    & MSE    & Mode (SD)    & MSE    & Mode (SD)    & MSE    \\ \midrule
GI                                                                     & $\beta_0$                   & 1.00                                                                  & 0.94 (0.06)  & 0.0034 & 0.94 (0.06)  & 0.0033 & 0.94 (0.06)  & 0.0038 & 0.40 (0.07)  & 0.3556 \\
                                                                       & $\beta_1$                   & -0.10                                                                 & -0.09 (0.05) & 0.0001 & -0.12 (0.07) & 0.0003 & -0.12 (0.06) & 0.0005 & -0.16 (0.08) & 0.0038 \\
                                                                       & $\rho_s$                    & 0.97                                                                  & 1.07 (0.52)  & 0.0110 & 1.04 (0.51)  & 0.0058 & 1.07 (0.52)  & 0.0101 & 0.99 (0.53)  & 0.0006 \\
                                                                       & $\rho_t$                    & 1.71                                                                  & 1.61 (0.80)  & 0.0117 & 1.57 (0.80)  & 0.0204 & 1.63 (0.80)  & 0.0072 & 1.47 (0.82)  & 0.0586 \\
                                                                       & $\rho_{st}$                 & 0.77                                                                  & 0.51 (0.39)  & 0.0655 & 0.53 (0.38)  & 0.0588 & 0.52 (0.39)  & 0.0627 & 0.61 (0.40)  & 0.0260 \\
                                                                       & $\nu$                       & 0.10                                                                  & 0.13 (0.03)  & 0.0009 & 0.81 (0.13)  & 0.5103 & 0.33 (0.06)  & 0.0510 & 0.11 (0.02)  & 0.0001 \\
                                                                       &                             &                                                                       &              &        &              &        &              &        &              &        \\
GSC                                                                    & $\beta_0$                   & 1.00                                                                  & 0.98 (0.04)  & 0.0005 & 0.98 (0.05)  & 0.0004 & 0.98 (0.04)  & 0.0005 & 0.77 (0.05)  & 0.0527 \\
                                                                       & $\beta_1$                   & -0.10                                                                 & -0.07 (0.04) & 0.0012 & -0.09 (0.05) & 0.0000 & -0.08 (0.05) & 0.0002 & -0.10 (0.05) & 0.0000 \\
                                                                       & $\rho_s$                    & 0.97                                                                  & 1.06 (0.62)  & 0.0087 & 1.07 (0.61)  & 0.0100 & 1.03 (0.61)  & 0.0039 & 0.99 (0.61)  & 0.0006 \\
                                                                       & $\rho_t$                    & 1.71                                                                  & 1.21 (1.07)  & 0.2552 & 1.21 (1.06)  & 0.2583 & 1.23 (1.05)  & 0.2384 & 1.09 (1.05)  & 0.3827 \\
                                                                       & $\rho_{st}$                 & 0.77                                                                  & 0.57 (0.46)  & 0.0400 & 0.54 (0.46)  & 0.0508 & 0.57 (0.46)  & 0.0412 & 0.65 (0.47)  & 0.0133 \\
                                                                       & $\nu$                       & 2.00                                                                  & 0.31 (0.72)  & 2.8471 & 2.08 (1.29)  & 0.0069 & 0.81 (1.33)  & 1.4156 & 0.29 (1.28)  & 2.9173 \\
                                                                       &                             &                                                                       &              &        &              &        &              &        &              &        \\
GSH                                                                    & $\beta_0$                   & 1.00                                                                  & 0.94 (0.06)  & 0.0032 & 0.94 (0.06)  & 0.0033 & 0.94 (0.06)  & 0.0035 & 0.41 (0.07)  & 0.3432 \\
                                                                       & $\beta_1$                   & -0.10                                                                 & -0.07 (0.05) & 0.0007 & -0.12 (0.07) & 0.0003 & -0.11 (0.06) & 0.0000 & -0.15 (0.08) & 0.0027 \\
                                                                       & $\rho_s$                    & 0.97                                                                  & 1.06 (0.52)  & 0.0092 & 1.06 (0.51)  & 0.0081 & 1.06 (0.52)  & 0.0097 & 1.02 (0.53)  & 0.0028 \\
                                                                       & $\rho_t$                    & 1.71                                                                  & 1.59 (0.80)  & 0.0157 & 1.56 (0.80)  & 0.0222 & 1.57 (0.80)  & 0.0219 & 1.47 (0.82)  & 0.0588 \\
                                                                       & $\rho_{st}$                 & 0.77                                                                  & 0.52 (0.38)  & 0.0640 & 0.53 (0.38)  & 0.0579 & 0.52 (0.38)  & 0.0641 & 0.61 (0.39)  & 0.0247 \\
                                                                       & $\nu$                       & 0.27                                                                  & 0.13 (0.03)  & 0.0195 & 0.83 (0.13)  & 0.3098 & 0.33 (0.06)  & 0.0035 & 0.11 (0.02)  & 0.0259 \\
                                                                       &                             &                                                                       &              &        &              &        &              &        &              &        \\
LN                                                                     & $\beta_0$                   & 1.00                                                                  & 1.26 (0.04)  & 0.0688 & 1.26 (0.04)  & 0.0687 & 1.26 (0.04)  & 0.0672 & 1.00 (0.04)  & 0.0000 \\
                                                                       & $\beta_1$                   & -0.10                                                                 & -0.05 (0.04) & 0.0021 & -0.10 (0.05) & 0.0000 & -0.08 (0.05) & 0.0004 & -0.10 (0.05) & 0.0000 \\
                                                                       & $\rho_s$                    & 0.97                                                                  & 1.01 (0.52)  & 0.0015 & 1.02 (0.52)  & 0.0028 & 0.99 (0.52)  & 0.0007 & 1.11 (0.54)  & 0.0211 \\
                                                                       & $\rho_t$                    & 1.71                                                                  & 1.45 (0.83)  & 0.0673 & 1.40 (0.83)  & 0.0961 & 1.37 (0.84)  & 0.1156 & 1.54 (0.90)  & 0.0304 \\
                                                                       & $\rho_{st}$                 & 0.77                                                                  & 0.59 (0.42)  & 0.0336 & 0.60 (0.42)  & 0.0287 & 0.61 (0.42)  & 0.0254 & 0.57 (0.43)  & 0.0391 \\
                                                                       & $\nu$                       & 0.27                                                                  & 0.15 (0.05)  & 0.0134 & 1.79 (0.35)  & 2.3014 & 0.52 (0.13)  & 0.0647 & 0.25 (0.14)  & 0.0002 \\ \bottomrule
\end{tabular}}
\end{table}


\section{Abundance of \textit{Nenia tridens} }
\label{sec:application}

Our research integrates several fundamental principles of ecology \citep{scheiner2008general} by exploring the bases of the heterogeneous distribution of organisms in space and time, and by linking such dynamics to the heterogeneous distribution of abiotic and biotic factors that represent local habitat characteristics. Indeed, this integration is a paramount challenge in ecology and biodiversity science and has critical ramifications for wildlife management and conservation action. Nonetheless, most ecological research considers spatio-temporal dynamics over periods of only $3$-$6$ years, thereby missing opportunities to consider long-term dynamics associated with long-term environmental variability. In contrast, we have taken advantage of long-term population data \citep{bloch2006context, willig1998long, willig2007cross} in a well-studied tropical ecosystem \citep{brokaw2012caribbean} that is subject climate-induced disturbances (i.e., cyclonic storms and droughts) to illustrate the utility of our new statistical model and evaluate the insights it provides for ecological understanding.

We investigated spatial, temporal and spatio-temporal trends in the abundance of {\it N. tridens} as well as in the environmental characteristics that may affect such variability. For this, we consider two possible fits. In one, we include only the basal effects of covariates, whereas in another we additionally allow regression coefficients to vary in time. The second approach was proposed to ascertain if any patterns arise when fitting temporal fixed effects for the covariates.

\begin{figure}
    \centering
    \includegraphics[width=0.90\textwidth]{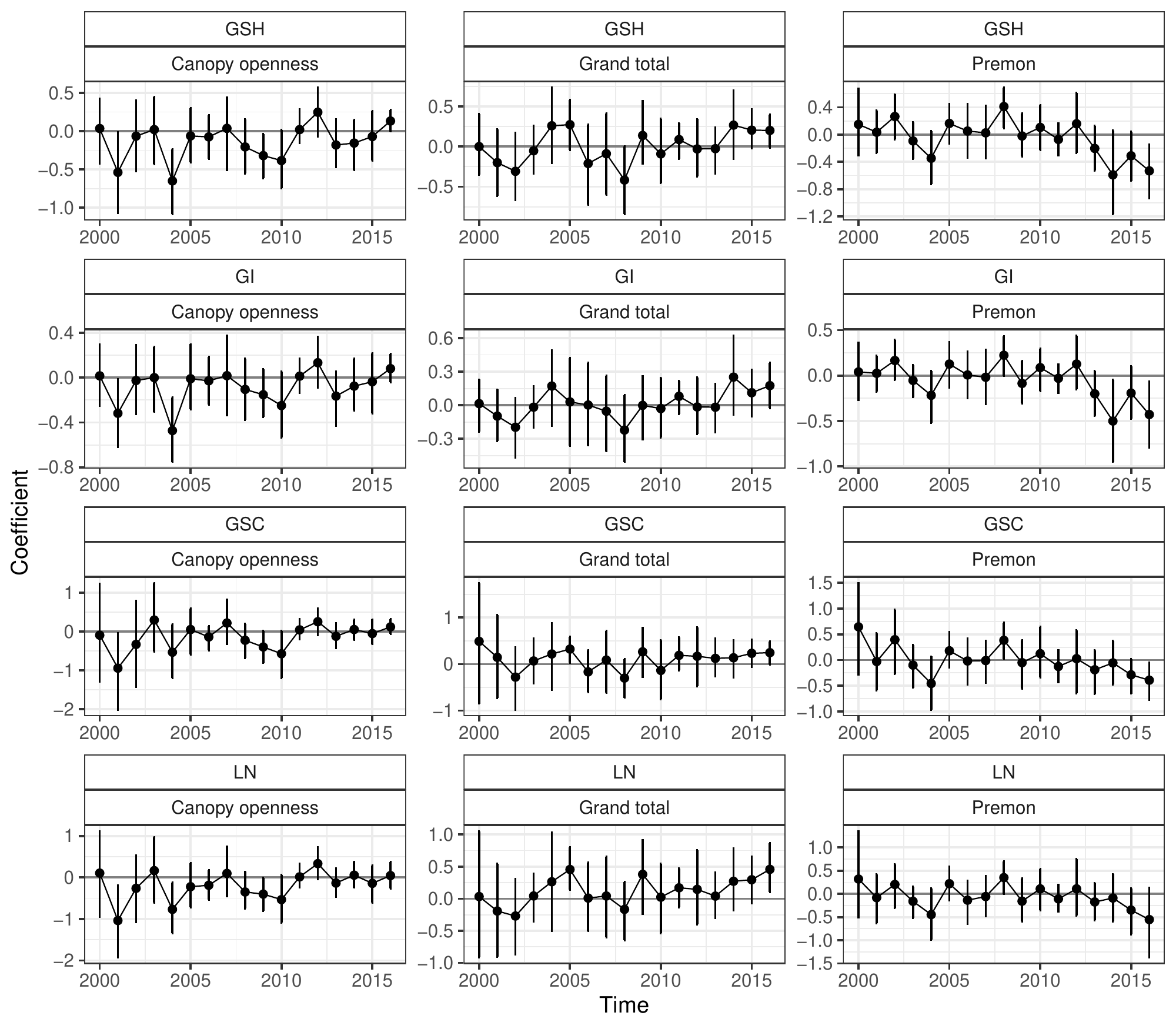}
    \caption{Time-varying coefficients.}
    \label{fig: beta tempo}
\end{figure}

As can be seen in Figure~\ref{fig: beta tempo} the evolution of the coefficients overtime does not suggest any pattern. Consequently, we believe that the constant fixed effect model is more parsimonious and should provide equivalent insights. 

Table~\ref{tab: parameters} shows parameter estimates for the model with constant fixed effects. As we can see the spatial and the temporal parameters were significantly greater than $0$, while the spatio-temporal dependence is not significantly different from $0$. Thus, for tabonuco forest, cross spatio-temporal dependence is not a significant factor affecting variation in abundance. In essence, the effects of space and time on the abundance of {\it N. tridens} are independent of each other.  As long as the fundamental niche of the species does not change over the time of the study (i.e. we are examining ecological rather than evolutionary dynamics), individuals should be responding to the same suite of environmental characteristics and should do so in the same manner over space and over time.  Previous research on {\it N. tridens} has shown that abundance is related to the same characteristics of the environment in two areas of tabonuco forest that differ from each other in the intensity of disturbance from Hurricane Hugo.  Although mean and variance of abundances differ greatly between the two regions, and the mean values for environmental characteristics are quite different between areas as well, the habitat characteristics that predict abundance did not differ significantly \citep{secrest1996legacy}.

To study the strength of the results obtained for $\rho_s$ and $\rho_t$, we compare the posterior estimates with the marginal limits, based on the diagonal dominance criterion, calculated for $\rho_s$ when $\rho_t = \rho_{st}=0$ and analogously for $\rho_t$. These limits are $\rho^{\max}_s = 2.25$ and $\rho^{\max}_t = 4.00$. This implies that $\hat \rho_s/\rho^{\max}_s \approx 0.76$ and $\hat \rho_t/\rho^{\max}_t \approx 0.85$, evidencing a strong association. Spatial dependence of abundance of a particular site about abundances at neighboring sites is likely due to the effect of immigration and emigration among those sites.  These lead to the greater similarity among sites in abundance than expected by chance.  Temporal dependence of abundance between consecutive time periods arises from the demographic process such as site-specific birth rates and death rates.
\begin{table}[]
\caption{Parameters estimation of the parameters for the application. Results are shown as Mode (Standard Deviation) and High Posterior Density (HPD) intervals.}
\label{tab: parameters}
\resizebox{.99\textwidth}{!}{%
\begin{tabular}{@{}lllllllll@{}}
\toprule
\multirow{2}{*}{Parameter} & GI           &                & GSC          &               & GSH          &                & LN           &                \\ \cmidrule(l){2-9} 
                           & Mode (SD)    & HPD 90\%       & Mode (SD)    & HPD 90\%      & Mode (SD)    & HPD 90\%       & Mode (SD)    & HPD 90\%       \\ \midrule
Intercept                  & 0.78 (0.09)  & (0.64, 0.91)   & 0.72 (0.11)  & (0.55, 0.89)  & 0.69 (0.12)  & (0.49, 0.85)   & -0.18 (0.11) & (-0.39, -0.03) \\
Elevation                  & 0.01 (0.06)  & (-0.09, 0.11)  & -0.03 (0.12) & (-0.21, 0.17) & 0.02 (0.10)  & (-0.14, 0.19)  & -0.04 (0.16) & (-0.29, 0.22)  \\
Slope                      & 0.04 (0.04)  & (-0.03, 0.10)  & 0.00 (0.06)  & (-0.10, 0.10) & 0.04 (0.06)  & (-0.04, 0.15)  & -0.02 (0.07) & (-0.13, 0.09)  \\
Grand total                & 0.04 (0.04)  & (-0.03, 0.11)  & 0.12 (0.06)  & (0.02, 0.23)  & 0.08 (0.06)  & (0.00, 0.18)   & 0.18 (0.07)  & (0.08, 0.30)   \\
Litter cover               &              &                &              &               &              &                &              &                \\
~~Low                        & ref.         &                & ref.         &               & ref.         &                & ref.         &                \\
~~Medium                     & 0.23 (0.08)  & (0.10, 0.35)   & 0.33 (0.13)  & (0.14, 0.55)  & 0.34 (0.13)  & (0.12, 0.55)   & 0.42 (0.14)  & (0.18, 0.66)   \\
~~High                       & 0.40 (0.11)  & (0.21, 0.56)   & 0.50 (0.16)  & (0.24, 0.72)  & 0.52 (0.16)  & (0.28, 0.76)   & 0.57 (0.17)  & (0.32, 0.86)   \\
Premon                     & -0.02 (0.04) & (-0.09, 0.05)  & -0.05 (0.06) & (-0.14, 0.07) & -0.03 (0.06) & (-0.13, 0.06)  & -0.07 (0.07) & (-0.19, 0.03)  \\
Canopy openness            & -0.02 (0.03) & (-0.08, 0.03)  & -0.01 (0.05) & (-0.09, 0.07) & -0.03 (0.05) & (-0.10, 0.05)  & -0.08 (0.06) & (-0.17, 0.01)  \\
$\rho_s$                   & 1.74 (0.16)  & (1.45, 1.92)   & 1.69 (0.20)  & (1.38, 1.94)  & 1.72 (0.15)  & (1.44, 1.90)   & 1.57 (0.18)  & (1.21, 1.79)   \\
$\rho_t$                   & 3.39 (0.15)  & (3.17, 3.63)   & 3.40 (0.16)  & (3.13, 3.64)  & 3.40 (0.15)  & (3.11, 3.60)   & 3.12 (0.18)  & (2.82, 3.38)   \\
$\rho_{st}$                  & -0.26 (0.14) & (-0.50, -0.06) & -0.25 (0.17) & (-0.51, 0.02) & -0.26 (0.14) & (-0.48, -0.04) & -0.03 (0.14) & (-0.25, 0.22)  \\
$\nu$                      & 0.07 (0.01)  & (0.06, 0.09)   & 0.51 (0.04)  & (0.44, 0.58)  & 0.20 (0.02)  & (0.16, 0.23)   & 0.09 (0.01)  & (0.07, 0.10)   \\ 
DIC     & 2189.73       &               & 2195.93          &              & {\bf 2188.92}       &                   & 2291.65       & \\
-2*LPML & 4496.43       &               & {\bf 3588.66}          &              & 3990.64       &                   & 3934.76 & \\
WAIC    & 2024.48       &               & 2026.46          &              & {\bf 2021.31}       &                   & 2115.74       & \\ \bottomrule
\end{tabular}}
\end{table}

Based on the model selection criteria, the GSH model was preferable since it has the best performance in 2 while GSC is preferable according to the LPML criterion. Thus, we can see that in this study the conventional log-normal approach does not provide the best fit. Grand total represents the total foliar volume of live vegetation in the understory of the forest, whereas litter cover estimates the volume of leaf litter on the forest floor.  Gastropods in general, and {\it N. tridens} in particular, use such live vegetation for the substrate on which to persist, or for food (the leaves themselves or algae, diatoms, or fungi that grow on them).  Leaf litter enhances humidity and decreases temperature on the forest floor.  Gastropods are very sensitive to desiccation, especially during periods of activity.  High humidity in the litter can mitigate microclimatic characteristics of the understory (e.g., during droughts or in tree fall gaps induced by cyclonic storms) that allow gastropods to persist and be active.  Moreover, leaf litter is a substrate on which micro-organisms grow that represent food sources.  Thus, the importance of these two characteristics is explicable in terms of the natural history of {\it N. tridens}, and bases on results of previous research \citep{secrest1996legacy}.

\section{Final remarks} \label{sec:final_remarks}

In this manuscript, an overview of many multivariate areal spatial models was considered and re-interpreted. Using the conditional mean and variance we show that parameter interpretation between the different literature proposals is not intuitive. With that in mind, a new valid multivariate spatial structure is introduced with an intuitive interpretation of parameters.

Such a multivariate structure is formulated in a spatio-temporal context and combined with the TGMRF approach. The TGMRF provides flexibility in the marginal distribution of the mean response and separates the mean structure from the dependence structure, thereby avoiding spatio-temporal confounding. As a by-product of this research, we provide the analyzed data and an \texttt{R} package (\url{https://github.com/douglasmesquita/TGMRF}) for this family of models, called TGMRF, for use by empiricists.

Spatio-temporal variation in counts of {\it N. tridens} is quite complex because of the environmental dynamics associated with disturbance and subsequent secondary succession in this tropical forest. Nonetheless, spatio-temporal interactions, after controlling for the spatial and temporal effects, do not provide additional predictive value. However, a strong positive spatial and temporal association is present. After controlling for the spatio-temporal dynamics, the density of vegetation in the understory and litter cover accounted for explaining the mean abundance at each site. 

Finally, as the model can be applied with regard to any hierarchical model as future studies we should include other likelihoods than the Poisson in the \texttt{R} package as well as other distribution families for the marginal. Models that can effectively ascertain the effects of space, time, and their interactions, all in the context of dynamically changing environmental characteristics, are critical tools for ecologists in the Anthropocene. Because the proposed approach and statistical tools are provided in \texttt{R}, these approaches should become widely adopted in a variety of ecological contexts and for any species of organism. At last, multivariate application over different species living in tabonuco forest can provide different insights about the complex dynamics of the ecological system.

\section*{Acknowledgments}
M. O. Prates acknowledges FAPEMIG and CNPq for partial financial support. Besides, this research was facilitated by grant numbers DEB-0218039, DEB-0620910, DEB-1239764, DEB-1546686, and DEB-1831952 from the National Science Foundation to the Institute of Tropical Ecosystem Studies, University of Puerto Rico, and the International Institute of Tropical Forestry as part of the Long-Term Ecological Research Program in the Luquillo Experimental Forest.  Additional support was provided by the USDA Forest Service, the University of Puerto Rico, the Department of Biological Sciences at Texas Tech University, and the Center for Environmental Sciences and Engineering at the University of Connecticut.  The staff of El Verde Field Station provided valuable logistical support in Puerto Rico. Finally, we thank the mid-sized army of students and colleagues who have assisted with the collection of field data over the years.

\bibliographystyle{chicago}
\bibliography{bibliografia}

\end{document}